\documentclass[pra, showpacs, twocolumn, floatfix]{revtex4}
\usepackage{graphicx}
\usepackage{amsmath, amsfonts, amssymb, bm}
\usepackage[]{psfrag}
\psfragscanoff 
\begin{document}
%%%%%%%%%%%%%%%%%%%%%%%%%%%%%%%%%%%%%%%%%%%%%%%%%%%%%%%%%%%%%%%%%%%%%%%%%%%%%%%%%%%%%%%%%%%%%%%%%%%%%%%%%%%%%%%%%%%%%%%%%%%%%

\title{Localization of atomic ensembles via superfluorescence}

\author{Mihai \surname{Macovei}}
\email{mihai.macovei@mpi-hd.mpg.de}
\affiliation{Max-Planck Institute for Nuclear Physics, 
Saupfercheckweg 1,D-69117 Heidelberg, Germany}

\author{J\"org \surname{Evers}}
\email{joerg.evers@mpi-hd.mpg.de}
\affiliation{Max-Planck Institute for Nuclear Physics, 
Saupfercheckweg 1,D-69117 Heidelberg, Germany}

\author{Christoph H. \surname{Keitel}}
\email{keitel@mpi-hd.mpg.de}
\affiliation{Max-Planck Institute for Nuclear Physics, 
Saupfercheckweg 1,D-69117 Heidelberg, Germany}

\author{M. Suhail \surname{Zubairy}}
\email{zubairy@physics.tamu.edu} 
\affiliation{Institute for Quantum Studies and Dept. of Physics,
Texas A\&M University, College Station, Texas 77843, USA}
\affiliation{Max-Planck Institute for Nuclear Physics, 
Saupfercheckweg 1,D-69117 Heidelberg, Germany}

\date{\today}

%%%%%%%%%%%%%%%%%%%%%%%%%%%%%%%%%%%%%%%%%%%%%%%%%%%%%%%%%%%%%%%%%%%%%%%%%%%%%%%%%%%%%%%%%%%%%%%%%%%%%%%%%%%%%%%%%%%%%%%%%%%%%
\begin{abstract}
The sub-wavelength localization of an ensemble of atoms 
concentrated to a small volume in space is investigated.
The localization relies on the interaction of the ensemble
with a standing wave laser field. The light scattered
in the interaction of standing wave field and atom
ensemble depends on the position of the ensemble relative
to the standing wave nodes. This relation can be described
by a fluorescence intensity profile, which depends on the standing
wave field parameters, the ensemble properties, and which
is modified due to collective effects in the ensemble of nearby 
particles. We demonstrate that the intensity profile can
be tailored to suit different localization setups.
Finally, we apply these results to two localization schemes.
First, we  show how to localize an ensemble fixed at a
certain position in the standing wave field. Second, we discuss
localization of an ensemble passing through the standing wave
field.
\end{abstract}
%%%%%%%%%%%%%%%%%%%%%%%%%%%%%%%%%%%%%%%%%%%%%%%%%%%%%%%%%%%%%%%%%
\maketitle
%%%%%%%%%%%%%%%%%%%%%%%%%%%%%%%%%%%%%%%%%%%%%%%%%%%%%%%%%%%%%%%%%%%%%%

\section{INTRODUCTION}
Nano-technology requires an accurate control of the interacting 
components, both in terms of detection and preparation.
This is a major motivation for the considerable attention that
was devoted recently to sub-wavelength localization of single 
particles. Several remarkable schemes 
were proposed to achieve this goal~\cite{QCW,TW,HL,KDR,LcZ,LPK,AK}. 
Of related interest is the problem of localizing and distinguishing two 
nearby particles. Optical resolution of two molecules at the 
nanometer scale and 
manipulation of their degree of entanglement was experimentally 
demonstrated in~\cite{He}, and the collective interaction between 
pairs of oriented nanostructures was considered in~\cite{Bar}. 
Also the measurement of the relative position of two atoms 
assisted by spontaneous emission~\cite{Su} or the measurements 
of interparticle separations on a scale smaller than 
the emission wavelength~\cite{JZ} were investigated in detail. 
Somewhat related, considerable effort is devoted to quantum 
lithography~\cite{Lt,hemmer}. The above localization schemes, 
however, have in common that they apply to the localization
of a single particle or to the measurement of relative position
of two individual particles. In many cases of interest,
however, an ensemble of particles is concentrated to a
small region in space such that the ensemble properties
become relevant, while the properties of the individual
ensemble constituents cannot be resolved or are rapidly 
fluctuating in time.

Therefore, here we describe a scheme capable of localizing an ensemble
of two-level atoms which are bunched together in a volume much 
smaller than an emission wavelength. 
Possible realizations include small clusters, few-atom 
impurities, or atoms trapped, e.g., in optical lattices.
The localization relies
on the coherent interaction with a standing-wave electromagnetic 
field.  Since the interatomic
distances are small, the atoms interact collectively via the 
environmental vacuum modes. One consequence of this is the
appearance of superfluorescence, i.e., the scattered light
intensity scales with the number of atoms $N$ squared 
($I \propto N^{2}$).
We find that the fluorescence light emitted collectively 
by the ensemble is a function of the ensemble position 
in the standing wave. In particular, for suitable standing
wave parameters and for ensemble positions
around the nodes of the standing wave field, the emitted
fluorescence intensity sharply drops to a minimum over a narrow
spatial region. The narrow width of the dip in the spatial
intensity profile is a direct consequence of the collectivity.
Since this collective fluorescence intensity profile
is our main observable, we discuss the profile in detail
in terms of the available free parameters, and show that
the profile can be tailored to suit a given localization
setup.
Based on these results, we then propose two schemes which exploit 
this spatial fluorescence intensity profile to localize an 
ensemble of atoms.
First, we assume the sample to be fixed within the standing wave 
field. In this case, the spatial fluorescence intensity profile can 
be scanned along the standing wave axis by changing the relative
phase of the laser fields forming the standing wave.
A continuous measurement of the intensity of the scattered
light throughout this scan reveals the position of the 
sample on a sub-wavelength scale. We further show that 
this setup also enables one
to measure the distance between two samples, the number
of atoms in a sample, or the linear dimension of the sample.
Second, we consider an atom cluster flying through the standing
wave field. Here a scanning is not possible due 
to the short interaction time of ensemble and standing wave
field. Rather, the absolute intensity of the scattered
light can be used to recover the crossing position of the
ensemble.

The article is organized as follows.
In Sec.~\ref{sec-theory}, we present our theoretical model 
for the ensemble interaction with the standing wave field
and derive the intensity profile as our main observable.
Sec.~\ref{sec-results} consists of three parts.
In the first part Sec.~\ref{sec-int}, we in detail study
the collective fluorescence profile numerically.
In Sec.~\ref{sec-sweep}, the results are applied to the
localization of an ensemble fixed in the standing wave field.
The last part Sec.~\ref{sec-single} discusses the localization
of an ensemble flying through the standing wave field.
Finally, Sec.~\ref{sec-summary} discusses and summarizes the
results.

%%%%%%%%%%%%%%%%%%%%%%%%%%%%%%%%%%%%%%%%%%%%%%%%%%%%%%%%
%%%%%%%%%%%%%%%%%%%%%%%%%%%%%%%%%%%%%%%%%%%%%%%%%%%%%%%%
%%%%%%%%%%%%%%%%%%%%%%%%%%%%%%%%%%%%%%%%%%%%%%%%%%%%%%%%
%%%%%%%%%%%%%%%%%%%%%%%%%%%%%%%%%%%%%%%%%%%%%%%%%%%%%%%%
\section{\label{sec-theory}THEORY}
In the usual mean-field, dipole, and rotating-wave approximations the interaction of such an atomic sample with 
an external laser field and the surrounding vacuum modes, in a frame rotating with the laser frequency $\omega_{L}$, 
is described by the Hamiltonian $H=H_{0} + H_{L} + H_{I}$ where 
%%%%%%%%%%%%%%%%%%%%%%%%%%%%%%%%%%%%%%%%%%%%%%%%%%%%%%%%%%%%%%%%%%%%%%%%%%
\begin{align} 
H_{0}&= \sum_{k}\hbar(\omega_{k}-\omega_{L})a^{\dagger}_{k}a_{k} + \sum^{N}_{j=1}\hbar(\omega_{0j}-\omega_{L})S_{zj}, 
\nonumber \allowdisplaybreaks[2] \\
H_{L}&=\sum^{N}_{j=1}\hbar\bigl( \Omega(\vec r_{j})S^{+}_{j} + \Omega^{\ast}(\vec r_{j})S^{-}_{j} \bigr), 
\nonumber \allowdisplaybreaks[2]\\
H_{I}&= i\sum_{k}\sum^{N}_{j=1}(\vec g_{k}\cdot \vec d_{j}) a^{\dagger}_{k}S^{-}_{j}e^{-i\vec k\cdot \vec r_{j}}
 + \textrm{ H.c.}\,. \label{Hm}
\end{align}
%%%%%%%%%%%%%%%%%%%%%%%%%%%%%%%%%%%%%%%%%%%%%%%%%%%%%%%%%%%%%%%%%%%%%%%%%%
Here $S^{\pm}_{j}$ are the raising and lowering operators for the $j$th atom, positioned at $\vec r_{j}$, and obeying the commutation 
relations $[S^{+}_{j},S^{-}_{l}]=2S_{zj}\delta_{jl}$, and $[S_{zj},S^{\pm}_{l}]=\pm S^{\pm}_{j}\delta_{jl}$ with $S_{zj}$
being the inversion operator. $a^{\dagger}$ and  $a$ are the radiation creation and annihilation operators satisfying 
the commutation relations $[a_{k},a^{\dagger}_{k^{'}}]=\delta_{kk^{'}}$, and $[a_{k},a_{k^{'}}]=[a^{\dagger}_{k},a^{\dagger}_{k^{'}}]=0$.

In Eq.~(\ref{Hm}), $H_{0}$ represents the free electromagnetic field (EMF) and free atomic Hamiltonians, respectively. 
The second term, i.e. $H_{L}$, describes the interaction of the atomic system with an external standing-wave coherent field. 
In general, the Rabi frequencies of the atoms in a standing wave are position-dependent since 
%%%%%%%%%%%%%%%%%%%%%%%%%%%%%%%%%%%%%%%%%%%%%%%%%%%%%%%%%%%%%%%%%%%%%%%%%%
\begin{eqnarray*}
\Omega(\vec r_{j})=\Omega_{j}\cos{(\vec k_{L}\cdot \vec r_{j})}, 
\end{eqnarray*}
%%%%%%%%%%%%%%%%%%%%%%%%%%%%%%%%%%%%%%%%%%%%%%%%%%%%%%%%%%%%%%%%%%%%%%%%%%
where $\Omega_{j} = (\vec d_{j} \cdot \vec E_{L})/\hbar$ while 
$E_{L}=|\vec E_{L}|$ is the amplitude of the electromagnetic field intensity with a wave vector $\vec k_{L}$ and 
$\vec d_{j}$ is the dipole moments of the atoms. 
Note that the scheme described here can also be generalized to
multiphoton transitions. Then, the Rabi frequency can be written as
\begin{equation}
\Omega(\vec r_{j})=\Omega_{j}^{(n)}\cos{(n\vec k_{L}\cdot \vec r_{j})}, 
\end{equation}
where $\Omega_{j}^{(n)}$ is a multiphoton Rabi-frequency arising
from an adiabatic elimination of intermediate states, and
$n$ denotes the number of photons involved in the multiphoton
process. In this way, the wavelength $\lambda=2\pi/k$ can be 
reduced to the effective wavelength $\lambda/n$, thus 
increasing the spatial resolution~\cite{hemmer}.
The last term in Eq.~(\ref{Hm}), $H_{I}$, takes into account the interaction 
of all atoms with the environmental vacuum modes. 

In the Born-Markov approximations the quantum dynamics of the driven multi-atom sample 
(each atom having identical transition frequency $\omega_{0}$) is governed by the master equation \cite{AGb,FS,KL,PR}:
%%%%%%%%%%%%%%%%%%%%%%%%%%%%%%%%%%%%%%%%%%%%%%%%%%%%%%%%%%%%%%%%%%%%%%%%%%
\begin{align}
\frac{d}{dt}\rho(t) &+ \frac{i}{\hbar}[\tilde H_{0},\rho] = \nonumber \\
&\quad \sum^{N}_{i,j=1}\{\gamma_{ij}(\omega_{0})[S^{+}_{j},S^{-}_{l}\rho] + \textrm{H.c.} \}\,, \label{ME}\\
\tilde H_{0}&=\hbar\sum_{j}[\Delta S_{zj}/2 + \Omega(\vec r_{j})S^{+}_{j} + \textrm{H.c.}]\,.
\end{align}
%%%%%%%%%%%%%%%%%%%%%%%%%%%%%%%%%%%%%%%%%%%%%%%%%%%%%%%%%%%%%%%%%%%%%%%%%%
Here, $\Delta = \omega_{0} - \omega_{L}$ is the detuning of atomic levels 
from the frequency of the driving field. Further,
%%%%%%%%%%%%%%%%%%%%%%%%%%%%%%%%%%%%%%%%%%%%%%%%%%%%%%%%%%%%%%%%%%%%%%%%%%%%%%%%%%%%%%%%%%%%%%%%%%%%%%%%%%%%%%%%%%%%%%%%%%%%
\begin{eqnarray}
\gamma_{jl}(\omega_{0})=\chi_{jl}(\omega_{0}) + i\Omega_{jl}(\omega_{0}), \label{CP}
\end{eqnarray} 
%%%%%%%%%%%%%%%%%%%%%%%%%%%%%%%%%%%%%%%%%%%%%%%%%%%%%%%%%%%%%%%%%%%%%%%%%%%%%%%%%%%%%%%%%%%%%%%%%%%%%%%%%%%%%%%%%%%%%%%%%%%%
where the collective parameters describing the mutual interactions among any atomic pair in the sample are given, respectively, 
by \cite{AGb,FS}
%%%%%%%%%%%%%%%%%%%%%%%%%%%%%%%%%%%%%%%%%%%%%%%%%%%%%%%%%%%%%%%%%%%%%%%%%%%%%%%%%%%%%%%%%%%%%%%%%%%%%%%%%%%%%%%%%%%%%%%%%%%%
\begin{eqnarray}
\chi_{jl}(\omega) &=& \frac{3\gamma}{2}\bigl \{[1-\cos^{2}{\xi_{jl}}]\frac{\sin(\omega r_{jl}/c)}{\omega r_{jl}/c} + [1-3\cos^{2}{\xi_{jl}}] \nonumber \\
&\times&\bigl [\frac{\cos(\omega r_{jl}/c)}{(\omega r_{jl}/c)^{2}}- \frac{\sin(\omega r_{jl}/c)}{(\omega r_{jl}/c)^{3}}\bigr ]\bigr \}, \nonumber \\
\Omega_{jl}(\omega) &=& \frac{3\gamma}{4}\bigl \{[\cos^{2}{\xi_{jl}}-1]\frac{\cos(\omega r_{jl}/c)}{\omega r_{jl}/c} + [1-3\cos^{2}{\xi_{jl}}] \nonumber \\
&\times& \bigl [\frac{\sin(\omega r_{jl}/c)}{(\omega r_{jl}/c)^{2}} + \frac{\cos(\omega r_{jl}/c)}{(\omega r_{jl}/c)^{3}}\bigr ]\bigr \}, 
\label{ColP}
\end{eqnarray}
%%%%%%%%%%%%%%%%%%%%%%%%%%%%%%%%%%%%%%%%%%%%%%%%%%%%%%%%%%%%%%%%%%%%%%%%%%%%%%%%%%%%%%%%%%%%%%%%%%%%%%%%%%%%%%%%%%%%%%%%%%%%
with $2\gamma = 4\omega^{3}_{0}d^{2}_{0}/(3\hbar c^{3})$ being the single-atom spontaneous decay rate. 
Here, we have assumed 
that all the dipole moments are identical and parallel, i.e. $d_{j}=d_{l} \cdots \equiv d_{0}$, and then $\xi_{jl}$ is the angle between the dipole moments $\vec d_{0}$ and $\vec r_{jl} = \vec r_{j} - \vec r_{l}$.

Inspecting the master equation (\ref{ME}) one can easily distinguish the part of it describing the coherent evolution of atoms
under the influence of the laser field, i.e. the term containing the Hamiltonian $\tilde H_{0}$, from that characterizing 
the collective spontaneous emission due to the vacuum modes, that is the terms proportional to $\rm{Re\{\gamma_{jl}(\omega_{0})\}}$.
The dipole-dipole interactions between the two-level atoms are described by the terms proportional to $\rm{Im\{\gamma_{jl}(\omega_{0})\}}$.
If the interparticle separations are small enough, that is $\omega r_{jl}/c \equiv k r_{jl} \to 0$ ($j \not = l$), then to second 
order in this parameter, Eqs.~(\ref{ColP}) reduce to 
%%%%%%%%%%%%%%%%%%%%%%%%%%%%%%%%%%%%%%%%%%%%%%%%%%%%%%%%%%%%%%%%%%%%%%%%%%
\begin{align}
\chi_{jl}(k) &= \gamma\{1 - \frac{1}{5}(k r_{jl})^{2}[1-\frac{1}{2}\cos^{2}\xi_{jl}]\}, \nonumber \\
\Omega_{jl}(k) &= 3\gamma\{[\cos^{2}\xi_{jl} -1](2/kr_{jl}-kr_{jl}) + [1- 
\nonumber \\
 3&\cos^{2}\xi_{jl}]  
[(kr_{jl})^{-1} - kr_{jl}/4 + 2(kr_{jl})^{-3}]\}/8. \label{apColP}  
\end{align}
%%%%%%%%%%%%%%%%%%%%%%%%%%%%%%%%%%%%%%%%%%%%%%%%%%%%%%%%%%%%%%%%%%%%%%%%%%
It is easy to realize that in this case $\chi_{jl} \to \gamma$ while $\Omega_{jl}$ reduces to the static dipole-dipole
interaction potential, i.e. 
\begin{eqnarray}
\Omega_{jl} = \frac{3\gamma}{4(kr_{jl})^{3}}\{1- 3\cos^{2}\xi_{jl}\}. \label{sdd}
\end{eqnarray}
%%%%%%%%%%%%%%%%%%%%%%%%%%%%%%%%%%%%%%%%%%%%%%%%%%%%%%%%%%%%%%%%%%%%%%%%%%
For lower atomic densities the collective parameters $\chi_{jl}$ and $\Omega_{jl}$ vanish because the atoms react independently from 
each other in this particular case. Finally, the master equation (\ref{ME}) describes adequately driven atomic samples of any shapes 
providing that retardation effects are negligible.

When dealing with smaller atomic systems of an arbitrary irregular shape it is hard to specify the orientation of dipole moments
relative to the interparticle separations. That is to say, we do not have a privileged angular distribution of photons as they
are emitted equally in all directions. Since there is no information on the dipole orientations, we average over all directions
the dipole-dipole interaction potential. Interestingly, the static dipole-dipole interaction given by Eq.~(\ref{sdd}) vanishes in 
this case. Then, according to the second term in Eq.~(\ref{apColP}), the averaged dipole-dipole interactions among the two-level 
emitters are given by the expression:
%%%%%%%%%%%%%%%%%%%%%%%%%%%%%%%%%%%%%%%%%%%%%%%%%%%%%%%%%%%%%%%%%%%%%%%%%%
\begin{eqnarray}
\Omega^{(av)}_{jl} = - \frac{\gamma}{2kr_{jl}}. \label{av_dd}
\end{eqnarray}
%%%%%%%%%%%%%%%%%%%%%%%%%%%%%%%%%%%%%%%%%%%%%%%%%%%%%%%%%%%%%%%%%%%%%%%%%%
Thus, the dipole-dipole potential reduces from a short-range to a long-range interaction, although the radiators
are close to each others. Moreover, the influence of the dipole-dipole interactions, in a two-atom system, was shown 
to be negligible in practice for interparticle separations such that $kr_{jl} \ge \pi/10$ \cite{LMe}. 

In what follows we shall apply Eqs.~(\ref{ME}-\ref{av_dd}) to the localization of a small atomic system within an
emission wavelength. Suppose that the linear dimension of the atomic sample is much less than the emission wavelength 
(say, for instance, smaller than $0.1\lambda$). Under this assumption the two-level emitters are almost in an equivalent
position relative to the driving standing-wave laser and we can omit the atomic indices from the expression characterizing 
the Rabi frequency, i.e $\Omega(r_{j}) \approx \Omega(r) \equiv \Omega(x)$. The master equations (\ref{ME}) transforms then 
into: 
%%%%%%%%%%%%%%%%%%%%%%%%%%%%%%%%%%%%%%%%%%%%%%%%%%%%%%%%%%%%%%%%%%%%%%%%%%
\begin{eqnarray}
\frac{d}{dt}\rho(t) &+& i[\tilde \Delta S_{z} + \Omega(x)(S^{+} + S^{-}) - \Omega_{d}S^{+}S^{-},\rho] \nonumber \\
&=& \gamma \{ [S^{+},S^{-}\rho] + [\rho S^{+},S^{-}] \}. 
\label{MER}
\end{eqnarray}
%%%%%%%%%%%%%%%%%%%%%%%%%%%%%%%%%%%%%%%%%%%%%%%%%%%%%%%%%%%%%%%%%%%%%%%%%%
Here $\tilde \Delta = \Delta + \Omega_{d}$ and $\{ S^{\pm}, S_{z} \}$ are collective atomic operators satisfying 
the standard commutation relations of su(2) algebra \cite{AGb,FS,KL,PR} while $\Omega_{d}$ is the dipole-dipole interaction 
potential considered identical for all radiators. Note that Eq.~(\ref{MER}) describes a small driven 
atomic system that involves symmetrized multiparticle states only.
The antisymmetric states are decoupled from the dynamics within our current framework,
and in the following we assume that they are not populated initially.
The dipole-dipole interaction $\Omega_{d}$ 
considerably shifts the symmetric states from the field resonance if the interparticle separations 
are very small~\cite{MFK}. Even though
the hypothesis of identical dipole-dipole interactions is, in general, not fulfilled, it admits to solve analytically 
the above master equation in the long-time limit. Thus, in order to get some insight on how the dipole-dipole 
interactions affect the localization processes of a small system as a whole we shall accept the hypothesis of 
identical dipole-dipole interactions between any pair in the sample.

The solving procedure of Eq.~(\ref{MER}) was described in \cite{KL,PR} for running wave lasers.
Adopting it to the case of a standing wave field, one arrives at the steady-state solution
%%%%%%%%%%%%%%%%%%%%%%%%%%%%%%%%%%%%%%%%%%%%%%%%%%%%%%%%%%%%%%%%%%%%%%%%%%
\begin{eqnarray}
\rho_{s} = Z^{-1}\sum^{N}_{n,m=0}C_{nm}(x)\bigl(S^{-}\bigr )^{n}\bigl(S^{+} \bigr)^{m}, \label{SSS}
\end{eqnarray}
%%%%%%%%%%%%%%%%%%%%%%%%%%%%%%%%%%%%%%%%%%%%%%%%%%%%%%%%%%%%%%%%%%%%%%%%%%
where 
\begin{eqnarray*}
C_{nm}(x) &=& (-1)^{n+m}\alpha^{-n}(\alpha^{\ast})^{-m}a_{nm}, \\
a_{nm} &=& \frac{\Gamma(1+n+\beta)\Gamma(1+m+\beta^{\ast})}{n!m!\Gamma(1+\beta)\Gamma(1+\beta^{\ast})},
\end{eqnarray*}
with  $\alpha=i\Omega(x)/(\gamma + i\Omega_{d})$ and $\beta=i\tilde \Delta/(\gamma + i\Omega_{d})$.
The normalization constant $Z$ is chosen such that $\rm{Tr\{\rho_{s}\}=1}$, i.e. 
\begin{eqnarray*}
Z=\sum^{N}_{n,m=0}(-1)^{n+m}\alpha^{-n}(\alpha^{\ast})^{-m}a_{nm}{\rm Tr}\{(S^{-})^{n}(S^{+})^{m}\}.
\end{eqnarray*}
%%%%%%%%%%%%%%%%%%%%%%%%%%%%%%%%%%%%%%%%%%%%%%%%%%%%%%%%%%%%%%%%%%%%%%%%%%

The trace can be performed using the following relations
%%%%%%%%%%%%%%%%%%%%%%%%%%%%%%%%%%%%%%%%%%%%%%%%%%%%%%%%%%%%%%%%%%%%%%%%%%
\begin{eqnarray}
S^{+}|s,l \rangle = \sqrt{(s-l)(s+l+1)}|s,l+1\rangle, \nonumber \\ 
S^{-}|s,l \rangle = \sqrt{(s+l)(s-l+1)}|s,l-1\rangle, \label{DiO}
\end{eqnarray}
%%%%%%%%%%%%%%%%%%%%%%%%%%%%%%%%%%%%%%%%%%%%%%%%%%%%%%%%%%%%%%%%%%%%%%%%%%
where the collective Dicke states $|s,l\rangle$, with $s=N/2$ and $-s \le l \le s$, are the eigenstates for the operator 
$S_{z}$ and the operator of the total "spin" $S^{2}$ \cite{PR}:
%%%%%%%%%%%%%%%%%%%%%%%%%%%%%%%%%%%%%%%%%%%%%%%%%%%%%%%%%%%%%%%%%%%%%%%%%%
\begin{eqnarray*}
S_{z}|s,l \rangle = l|s,l\rangle, \\ 
S^{2}|s,l \rangle = s(s+1)|s,l\rangle.
\end{eqnarray*}
%%%%%%%%%%%%%%%%%%%%%%%%%%%%%%%%%%%%%%%%%%%%%%%%%%%%%%%%%%%%%%%%%%%%%%%%%%
Thus, 
\begin{eqnarray}
&{}&{\rm Tr}\{(S^{-})^{n}(S^{+})^{m}\}=\sum^{s}_{l=-s}\langle l,s|(S^{-})^{n}(S^{+})^{m}|l,s\rangle \nonumber \\
&=& \delta_{n,m}\sum^{s}_{l=-s}\langle l,s|(S^{-})^{n}(S^{+})^{n}|l,s\rangle, \label{sp} 
\end{eqnarray}
and, then 
\begin{eqnarray}
Z=\sum^{N}_{n=0}a_{nn}|\alpha|^{-2n}\frac{(N+n+1)!(n!)^{2}}{(N-n)!(2n+1)!}. \label{Z}
\end{eqnarray}
%%%%%%%%%%%%%%%%%%%%%%%%%%%%%%%%%%%%%%%%%%%%%%%%%%%%%%%%%%%%%%%%%%%%%%%%%%

The intensity of the collective resonance fluorescence emitted by driving the multiparticle system is calculated 
taking into account that this quantity is proportional to the first order atomic correlator, i.e. 
$I \propto \langle S^{+}S^{-}\rangle$. Then, using Eq.~(\ref{SSS} - \ref{Z}), one obtains:
%%%%%%%%%%%%%%%%%%%%%%%%%%%%%%%%%%%%%%%%%%%%%%%%%%%%%%%%%%%%%%%%%%%%%%%%%%
\begin{eqnarray}
I(x) = Z^{-1}\sum^{N}_{k=1}C_{k-1k-1}(x)\frac{(N+k+1)!(k!)^{2}}{(N-k)!(2k+1)!}. \label{It}
\end{eqnarray}
%%%%%%%%%%%%%%%%%%%%%%%%%%%%%%%%%%%%%%%%%%%%%%%%%%%%%%%%%%%%%%%%%%%%%%%%%%

\section{\label{sec-results}RESULTS}
We now turn to the discussion of Eq.~(\ref{It}).
First, we study the dependence of the collective fluorescence intensity
on the various external parameters. Second, we show how the
collective fluorescence intensity can be used to precisely locate
a sample of particles which is fixed in space inside the standing
wave field. Third, we discuss the localization of a collection 
of particles flying through the cavity field.

%%%%%%%%%%%%%%%%%%%%%%%%%%%%%%%%%%%%%%%%%%%%%%%%%%%%%%%%%%%%%%%%%%%%%%%%%%%  
\begin{figure}[t]
\includegraphics[height=4.5cm]{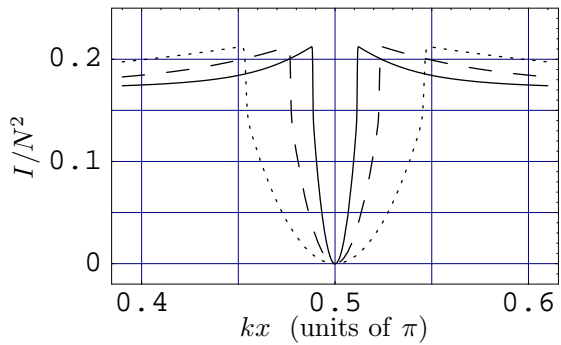}
\caption{\label{fig-1} The dependence of the collective steady-state resonance fluorescence intensity 
$I/N^{2}$ as function of $kx$. The solid, dashed and dotted curves are for $\Omega/(N\gamma)=100; 50$ and $25$, respectively. 
Other parameters are: $\Omega_{d}/\gamma=-10$,  $\Delta/(N\gamma)=0.5$ and $N=100$.}
\end{figure}
%%%%%%%%%%%%%%%%%%%%%%%%%%%%%%%%%%%%%%%%%%%%%%%%%%%%%%%%%%%%%%%%%%%%%%%%%%%  

\subsection{\label{sec-int}Collective fluorescence intensity}

Fig.~\ref{fig-1} shows the collective fluorescence intensity 
versus the position of the multiparticle collection in the
standing wave field. Note that $kx=\pi/2$ corresponds to
a node of the standing wave field, and thus the 
fluorescence intensity vanishes for particles located at 
this point in space. The parameters in Fig.~\ref{fig-1}
are number of atoms $N=100$, dipole-dipole coupling constant
$\Omega_{d}/\gamma=-10$, and detuning $\Delta/(N\gamma)=0.5$.
The solid line shows a standing wave Rabi frequency
$\Omega/(N\gamma)=100$, the dashed line is for 
$\Omega/(N\gamma)=50$, and the dotted one is for 
$\Omega/(N\gamma)=25$. It can be seen that with decreasing 
Rabi frequency, the width of the dip in the fluorescence
intensity around the node at $kx=\pi/2$ increases. Thus
a strong driving field allows for a narrow region in space
that leads to vanishing fluorescence intensity, while a
weaker field gives rise to fluorescence intensity over
a wider range of positions.

%
%%%%%%%%%%%%%%%%%%%%%%%%%%%%%%%%%%%%%%%%%%%%%%%%%%%%%%%%%%%%%%%%%%%%%%%%%%%  
\begin{figure}[t]
\includegraphics[width=8cm]{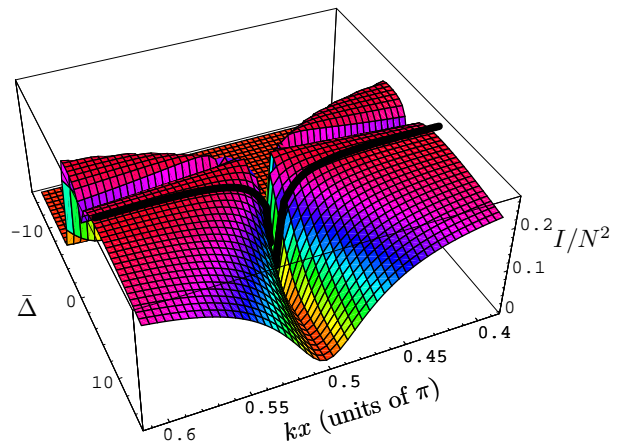}
\caption{\label{fig-2}(Color online) 
The steady-state collective resonance fluorescence intensity $I/N^{2}$ 
as function of $kx$  and $\bar \Delta = \Delta/(N\gamma)$. Here 
$N=100$, $\Omega/(N\gamma)=50$
and $\Omega_{d}/\gamma = -10$.
The black line on top of the surface plot indicates the position
$\bar \Delta = |\Omega_d|/\gamma = 10$.}
\end{figure}
%%%%%%%%%%%%%%%%%%%%%%%%%%%%%%%%%%%%%%%%%%%%%%%%%%%%%%%%%%%%%%%%%%%%%%%%%%  
%%%%%%%%%%%%%%%%%%%%%%%%%%%%%%%%%%%%%%%%%%%%%%%%%%%%%%%%%%%%%%%%%%%%%%%%%%%  
\begin{figure}[t]
\includegraphics[width=8cm]{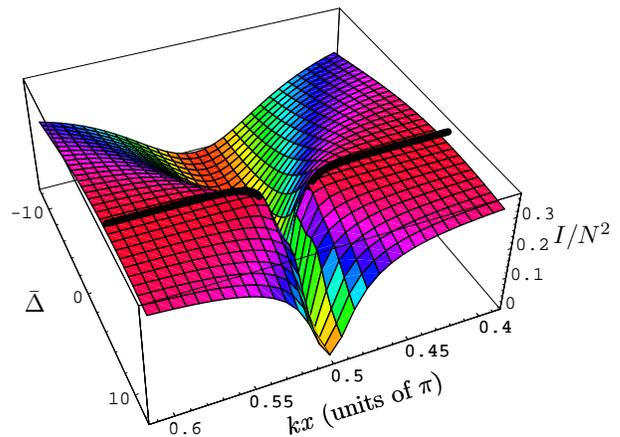}
\caption{\label{fig-3}(Color online) The steady-state collective resonance fluorescence intensity $I/N^{2}$ as function 
of $kx$ 
 and $\bar \Delta = \Delta/(N\gamma)$. Here $N=2$, $\Omega/(N\gamma)=50$
and $\Omega_{d}/\gamma = -5$. The black line on top of the surface 
plot indicates the position
$\bar \Delta = 5$ around which the two-atom symmetric collective state
is located.}
\end{figure}
%%%%%%%%%%%%%%%%%%%%%%%%%%%%%%%%%%%%%%%%%%%%%%%%%%%%%%%%%%%%%%%%%%%%%%%%%%  

The second free parameter is the detuning $\Delta$ between the
driving field frequency and the bare transition frequency 
$\omega_0$ of the individual atoms in the sample. 
Fig.~\ref{fig-2} shows the dependence of the collective fluorescence
intensity versus the position of the sample on this detuning.
Note that the y-axis of this figure is a scaled detuning 
$\bar\Delta = \Delta/(N\gamma)$. Thus, $\bar\Delta=0$
corresponds to the resonance case $\Delta = 0$, whereas
$\bar\Delta > 0$ indicates $\Delta>0$ and thus 
$\omega_L < \omega_0$. A qualitative understanding of this figure
can be gained from the case of a two atom sample,
as shown in Fig.~\ref{fig-3}. In a collective state
basis, the two-atom sample corresponds to a collective ground state
$|g_a, g_b\rangle$ at energy 0, where each of the two atoms 
$a,b$ is in its respective ground state,
an excited collective state $|e_a, e_b\rangle$
at energy $2\hbar \omega_0$, and
a symmetric [antisymmetric] collective state at energy 
$\hbar(\omega_0 + \Omega_d)$ [$\hbar(\omega_0 - \Omega_d)$]:
\begin{subequations}
\begin{eqnarray}
|S\rangle = \frac{1}{\sqrt{2}}\left (|g_a, e_b\rangle  + |e_a, g_b\rangle \right ) \,,\\
|A\rangle = \frac{1}{\sqrt{2}}\left (|g_a, e_b\rangle  - |e_a, g_b\rangle \right ) \,.
\end{eqnarray}
\end{subequations}
In the limit
of small interparticle distance chosen in our analysis,
the asymmetric state decouples from the dynamics,
such that we are essentially left with a three-state
ladder system. In Figure~\ref{fig-3}, the symmetric state
is located at $\bar \Delta = 5$, since the dipole-dipole
coupling is chosen as $\Omega_d = -5\gamma$. 
%
%
%%%%%%%%%%%%%%%%%%%%%%%%%%%%%%%%%%%%%%%%%%%%%%%%%%%%%%%%%%%%%%%%%%%%%%%%%%%  
\begin{figure}[t]
\includegraphics[width=8cm]{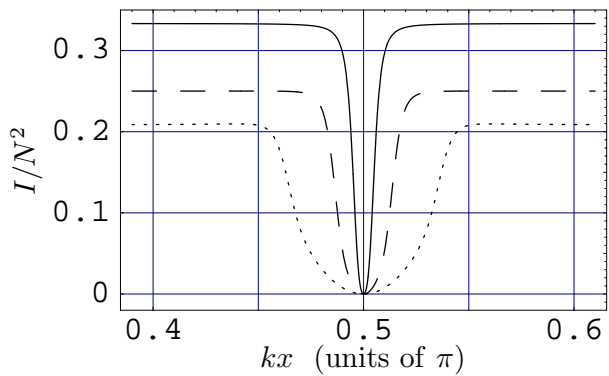}
\caption{\label{fig-4}Collective fluorescence versus atom ensemble
position in the standing wave field for different sizes of the
ensemble. The parameters are $\Omega_{d}/\gamma = -5$,
$\Delta/\gamma=10$, and $\Omega/\gamma=100$.
The solid line is for number of atoms $N = 2$, the dashed shows
the case $N = 4$, and the dotted is for $N=8$.}
\end{figure}
%%%%%%%%%%%%%%%%%%%%%%%%%%%%%%%%%%%%%%%%%%%%%%%%%%%%%%%%%%%%%%%%%%%%%%%%%%  
%
It can be seen that the width of the intensity dip is minimal
if the driving field frequency is in resonance with the
symmetric state, and becomes wider in moving away from the
resonance. 
No additional structure is visible around $\bar \Delta = -5$, 
where the asymmetric state is located, since it is
decoupled.
Note that for vanishing dipole-dipole interaction, $\Omega_d=0$, 
Figs.~\ref{fig-2} and \ref{fig-3} would exhibit features symmetric with
respect to both the planes given by $\bar{\Delta} = 0$
and $kx=\pi/2$.

These results from the two-atom case directly carry over to the
many-particle sample. The minimum width of the intensity
dip is close to the position where symmetric state combinations
can be expected, whereas no structure can be found 
towards asymmetric collective states.
A direct identification of the position of the detuning with
minimum dip width is difficult, however, since the collective-state
basis of a multi-particle sample includes many symmetric collective
states. In the example of Fig.~\ref{fig-2}, in a very small range
around the node $kx=\pi/2$ the dip at $\bar\Delta=10$ is narrowest,
but its width increases faster in moving away from the node
than it does for slightly lower values of $\bar\Delta$.
The many-particle
case also shows an additional structure at $\bar\Delta = 0$,
see Fig.~\ref{fig-2}, which, however, is not of interest for
our current localization scheme.

Finally, in Fig.~\ref{fig-4}, we show the dependence of the 
collective intensity on the number of particles in the sample.
This is different from the previous results, since by changing
the number of atoms, both the scaled Rabi frequency
$\Omega/(N\gamma)$ and the scaled detuning $\Delta/(N\gamma)$
are changed at the same time. It can be seen from Fig.~\ref{fig-4}
that for a given standing wave intensity and a given detuning,
varying the number of atoms in the ensemble changes the width
of the intensity dip at the nodes. This can be understood by
noting that a change of the number of atoms effectively shifts
the position of the symmetric state resonance. Since the
laser field frequencies are kept fixed in Fig.~\ref{fig-4}, 
this corresponds to moving along the $\bar\Delta$ axis
in Figs.~\ref{fig-2} and \ref{fig-3}. Therefore, different
widths of the intensity dip can be observed.
The maximum intensity changes with $N$ in Fig.~\ref{fig-4},
since in this figure the unscaled Rabi frequency $\Omega$ is 
kept fixed.
It should be noted that the parameters in Fig.~\ref{fig-4} are such 
that for any shown number of atoms, the scaled Rabi frequency
$\Omega/(N\gamma)$ dominates the dynamics. A further increase
of the number of atoms which leads to $|\Omega_d|/\gamma \gg \Omega/(N\gamma)$
shifts the relevant collective states out of the laser field
resonance such that the total fluorescence intensity vanishes.

%%%%%%%%%%%%%%%%%%%%%%%%%%%%%%%%%%%%%%%%%%%%%%%%%%%%%%%%%%%%%%%%%%%%%%%%%%%  
\begin{figure}[t]
\includegraphics[width=6cm]{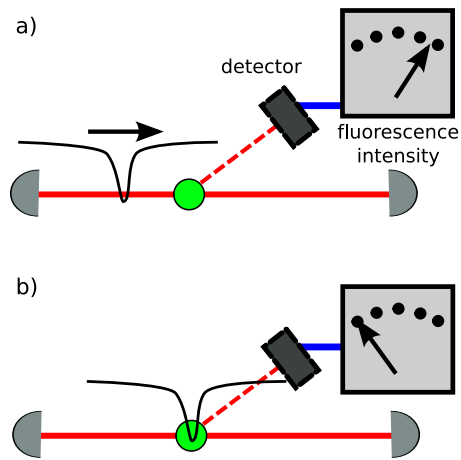}
\caption{\label{fig-sweep}(Color online) 
Scanning-dip scheme. The figure depicts
a possible experimental implementation of our scheme. An atom ensemble
(green dot) is assumed fixed inside the standing wave field. A detector measures
the scattered fluorescence light, while the phase of the standing wave
field is varied. The black curve symbolizes the fluorescence intensity
profile as, e.g., shown in Fig.~\ref{fig-1}. If the intensity dip does
not coincide with the ensemble position, then a high intensity of 
fluorescence is detected.
But if the dip sweeps across the ensemble position, then the measured
intensity drops to a minimum over a narrow spatial range, thus providing a 
sub-wavelength localization.}
\end{figure}
%%%%%%%%%%%%%%%%%%%%%%%%%%%%%%%%%%%%%%%%%%%%%%%%%%%%%%%%%%%%%%%%%%%%%%%%%%  

\subsection{\label{sec-sweep}Scanning-dip spectroscopy}
After the discussion of the collective fluorescence intensity
as our main observable, we now turn to the application of this
observable to the localization of a collection of atoms.
As the first setup, we consider a collection of atoms which is
fixed inside the standing wave field at an unknown position.
In order to detect the position of the sample, the total 
collective fluorescence intensity is continuously
monitored, which may already provide a coarse position measurement. 
Then the relative phase of the two counter-propagating
fields forming the standing wave is changed, such that the
nodal structure of the field shifts along the standing wave
propagation axis. Throughout this shift, the detected intensity
is modulated in time with the collective fluorescence 
intensity profile, as depicted in Fig.~\ref{fig-sweep}. 
If the intensity profile is located such that its dip does
not coincide with the actual position of the sample, then the 
intensity is near its maximum value, see Fig.~\ref{fig-sweep}(a).
But if the two positions coincide, then
the intensity vanishes, see Fig.~\ref{fig-sweep}(b).
Obviously, for this scheme it is desirable to have the intensity
dip as narrow as possible in order to achieve a localization
well below the usual diffraction limit. According to our results
of Sec.~\ref{sec-int}, this can be achieved by using a
strong standing wave field and by tuning it close to the
symmetric collective state resonance. For instance, the
solid curve in Fig.~\ref{fig-1} has a width of about
$\Delta(kx) = 0.02\pi$, corresponding to
$\Delta x = 0.01 \lambda$. Note, however, that the obtained
accuracy is also limited by the spatial size of the ensemble,
if the dip width is smaller than the linear dimension of the
sample.

%%%%%%%%%%%%%%%%%%%%%%%%%%%%%%%%%%%%%%%%%%%%%%%%%%%%%%%%%%%%%%%%%%%%%%%%%%%  
\begin{figure}[t]
\includegraphics[width=6cm]{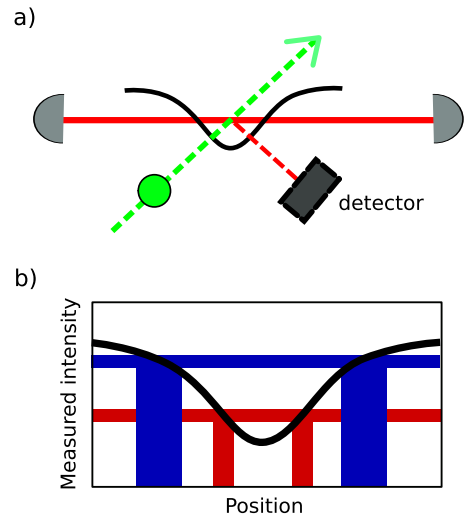}
\caption{\label{fig-single}(Color online) 
Single-pass localization. (a) shows the schematic setup. An ensemble
indicated by the green dot passes through a standing wave field.
The intensity of the scattered fluorescence light is measured.  
(b) Via the fluorescence intensity profile, the measured intensity
can be related to the sub-wavelength particle position. The higher
the slope of the intensity profile, the better is the accuracy of
the localization. In this figure, the horizontal bars indicate 
two measurement outcomes with a certain uncertainty. The vertical
bars show the corresponding potential positions. The black curve is
a fluorescence intensity profile as, e.g., shown in Fig.~\ref{fig-1}.}
\end{figure}
%%%%%%%%%%%%%%%%%%%%%%%%%%%%%%%%%%%%%%%%%%%%%%%%%%%%%%%%%%%%%%%%%%%%%%%%%%  

The present scheme can also be used to measure the distance 
between the center of masses of two ensembles by relating the required phase 
shifts between two positions with vanishing intensity to 
the measured intensity profile.
Further, since the width of the intensity dip depends on the number
of atoms in the ensemble if all other parameters remain fixed 
[see Fig.~\ref{fig-4}], this scheme can also be used to obtain
the number of atoms in the ensemble located in the standing wave field
by measuring the width of the intensity dip via the sweep of the
standing wave phase.
Finally, the spatial dimension of the sample can be measured from
the width of the intensity dip if the other parameters are known.
It should be noted that, in contrast to the precision position detection,
 these measurements relying on the
determination of the width of the intensity dip are relative
measurements in the sense that they do not require a reference 
to obtain the absolute standing wave phase. Only the change
of the phase is relevant, and thus also knowledge of the 
actual position of the sample within the standing wave field 
is not required for these relative measurements. 
The maximum attenuation of the fluorescence intensity  depends
on the  width of the collection relative to the width of the dip in the
intensity profile. If the profile dip is narrower than the 
sample, then the dip only affects part of the sample. Still,
this will result in a sudden reduction of the fluorescence
intensity, which is sufficient to determine the onset of the
overlap of intensity dip and atom sample. Also, a continuous
scan of the standing wave phase results in repeated intensity
dips over time which can be used to suppress statistical errors
in the measurement. 

\subsection{\label{sec-single}Single-pass localization}
In this section, we discuss a different experimental setup.
We now assume that a collection of atoms (atomic cluster) 
flies through a standing
wave field at an unknown position on the standing wave field axis.
The aim is to gain as much information on the position as possible
by measuring the collective resonance fluorescence. Obviously,
the scanning-dip scheme described in Sec.~\ref{sec-sweep} is not
suitable for this kind of setup, since the change of the standing
wave phase is too slow as compared to the interaction time
of field and atom ensemble. In the present scheme, we only
have to require that the time of flight $\tau_f$
through the standing
wave field is much larger than the time $\tau_s$ 
needed to evolve into the steady state. 
For example, for a thermal beam with velocity $300$~m/s and
a standing wave field width of $1$~mm, the flight time is
about $\tau_f = 3 \cdot 10^{-6}$~s. The steady-state time $\tau_s$ is of the order 
$(N\gamma)^{-1}$. For $\gamma = 10$~MHz and $N=10$, one obtains
$\tau_s = 10^{-8}$~s and thus $\tau_s \ll \tau_f$.
Note that the preparation of atom clusters has been discussed in~\cite{cluster}.

We now make use of the fact that position information can be gained from 
the absolute value of the scattered light intensity during the
flight of the ensemble through the field.
The schematic setup is shown in Fig.~\ref{fig-single}(a).
The measured intensity allows to fix a horizontal section in 
a collective fluorescence intensity plot versus ensemble position 
as shown in Fig.~\ref{fig-single}(b).
Ideally, this section provides a set of few discrete points
where the intensity profile crosses the measured intensity.
These points correspond to the potential positions of the
ensemble. From the two examples in Fig.~\ref{fig-single}(b)
it is clear that the localization for a given intensity measurement
with a measurement uncertainty becomes better with increasing slope
of the intensity profile. A high slope, however, leads to 
a pronounced plateau with almost constant intensity in between the
dips. This means that, in case of a narrow dip, for large parts of 
the single wavelength width only a rather inaccurate localization 
is possible. Therefore, in contrast to the sweep scheme in 
Sec.~\ref{sec-sweep}, in this setup a wide intensity dip 
is desirable in order to achieve sub-wavelength localization for all 
possible positions.
The reason is that for a wide intensity dip, wide plateaus in the 
intensity profile are avoided.
A wide intensity dip can be achieved,  for example,
by working with weaker standing wave fields or far away
from the symmetric collective state resonance.

%%%%%%%%%%%%%%%%%%%%%%%%%%%%%%%%%%%%%%%%%%%%%%%%%%%%%%%%
%%%%%%%%%%%%%%%%%%%%%%%%%%%%%%%%%%%%%%%%%%%%%%%%%%%%%%%%
%%%%%%%%%%%%%%%%%%%%%%%%%%%%%%%%%%%%%%%%%%%%%%%%%%%%%%%%
%%%%%%%%%%%%%%%%%%%%%%%%%%%%%%%%%%%%%%%%%%%%%%%%%%%%%%%%
\section{\label{sec-summary}DISCUSSION AND SUMMARY}
We have described a scheme to localize small atomic 
samples with sub-wavelength accuracy. The scheme relies on 
measuring the super-fluorescence radiation scattered in a 
standing wave field. We have demonstrated that external parameters 
such as the strength of the applied lasers or the detuning 
from the atomic resonance are convenient tools to tailor the 
localization region for a given experimental setup.
Based on these results, two possible experimental situations have been
considered. First, for fixed samples, a scanning-dip spectroscopy
was proposed. Here, the standing wave field phase is changed
in order to scan the fluorescence intensity profile along the
cavity axis in order to reveal the actual position of the sample.
This setup also allows for a number of relative measurements,
for example, of distance between two collections, of the number
of atoms in a sample, or of the linear dimension of the sample.
Second, for samples passing through the standing wave
field only once, a single-pass scheme was discussed, which relates
the maximum intensity measured to the passing position 
of the sample. Our scheme can be generalized to the multi-photon
case.

%%%%%%%%%%%%%%%%%%%%%%%%%%%%%%%%%%%%%%%%%%%%%%%%%%%%%%%%%%%%%%%%%%%%%%%%%%%%%%%%%%%%%%%
{\small $^\ast$ On leave from \it{Technical University of Moldova, Physics Department, 
\c{S}tefan Cel Mare Av. 168, MD-2004 Chi\c{s}in\u{a}u, Moldova.}}
%%%%%%%%%%%%%%%%%%%%%%%%%%%%%%%%%%%%%%%%%%%%%%%%%%%%%%%%%%%%%%%%%%%%%%%%%%%%%%%%%%%%%%%

%%%%%%%%%%%%%%%%%%%%%%%%%%%%%%%%%%%%%%%%%%%%%%%%%%%%%%%%%%%%%%%%%%%%%%%%%%%%%%%%%%%%%%%

\end{document}